\documentclass[letterpaper,11pt]{article}
\usepackage{tabularx} 
\usepackage{amsmath}  
\usepackage{amsfonts}
\usepackage{graphicx} 
\usepackage[margin=1.25in,letterpaper]{geometry} 
\usepackage{cite} 
\usepackage[final]{hyperref} 
\hypersetup{
	colorlinks=true,       
	linkcolor=[rgb]{0.55,0,0.5},        
	citecolor=[rgb]{0,0,0.6},        
	filecolor=magenta,     
	urlcolor=blue         
}

\newcommand{\opunit}{\text{1}\kern-0.22em\text{l}}

\DeclareMathAlphabet{\mathpzc}{OT1}{pzc}{m}{it}

\newcommand{\id}{\textrm{d}}

\newcommand{\JR}{{\mathbb R}}
\newcommand{\ee}{\end{equation}}
\newcommand{\bea}{\begin{eqnarray}}
\newcommand{\eea}{\end{eqnarray}}

\newcommand{\nn}{\nonumber }

\begin{document}

\title{The induced motion of a probe coupled to a bath\\ with random resettings}

\author{Christian Maes and Thimoth\'ee Thiery\\ {\it Instituut voor Theoretische Fysica, KU Leuven}}


\maketitle

\begin{abstract} We consider a probe linearly coupled to the center of mass of a nonequilibrium bath.  We study the induced motion on the probe for a model where a resetting mechanism is added to an overdamped bath dynamics with quadratic potentials.  The fact that each bath-particle is at random times being reset to a fixed position is known for optimizing diffusive search strategies, but here stands for the nonequilibrium aspect of the bath. In the large bath scaling limit the probe is governed by an effective Langevin equation. Depending on the value of the parameters, there appear three regimes: (i) an equilibrium-like regime but with a reduced friction and an increased effective temperature; (ii) a regime where the noise felt by the probe is continuous but nonGaussian and exhibits fat-tails; (iii) a regime with a nonGaussian noise exhibiting power-law distributed jumps. The model thus represents an exactly solvable case for the origin of nonequilibrium probe dynamics. 
\end{abstract}


\section{Introduction}
An interesting recent development in nonequilibrium statistical mechanics is to study the motion of ``probes'' in contact with a nonequilibrium environment \cite{simi,jsp,stef,carlo}.  The probe can be a colloid, a real additional particle immersed in the bath or, more abstractly, the probe can also represent a macroscopic variable which is moving on a slower time-scale than the relevant bath degrees of motion.   The bath consists of a huge number of particles that interact with the probe via some macroscopic interaction, i.e., involving many particles.  While the problem is thus formally identical with deriving Brownian motion or Langevin dynamics \cite{vK,bha}, in the present study the bath is out of thermal equilibrium (even before the probe disturbs it).   The bath-particles are driven and dissipate energy in yet another (now equilibrium) environment at fixed temperature.  The question is to find the relevant changes in systematic and fluctuating forces on the probe due to the nonequilibrium condition of the bath.\\
  There are many realizations of such systems having three levels (probe, nonequilibrium medium and thermal environment) of description, but much remains to be explored when the medium is out-of-equilibrium.  Other similar studies of the induced motion on a probe in contact with a nonequilibrium medium include \cite{simi,jsp,stef,carlo}.  Here we take a bath dynamics where the nonequilibrium is due to random resettings of the particles at a position $A$. Stochastic resetting  \cite{ss} has recently been studied in connection with algorithmic searches \cite{ma2,ma1,ma3,ma4,rold,pal,pal2} or as an elementary solvable example of a system with an out-of-equilibrium steady state \cite{seifert}. Here the resetting stands for the non-equilibrium aspect of the bath and we think of our model as being similar to having a bath in a randomly pulsating volume, or a gas that is maintained out-of-equilibrium by random kicks. The technical advantage of the present set-up is that between any two resettings each particle undergoes an undriven motion in contact with the probe.  We can calculate basically everything exactly for a linear version of the bath with resettings, which thus represents a useful reference case for further explorations. Despite its simplicity, we find that the model exhibits an interesting phenomenology. In a first regime we find that the resetting induces an equilibrium-like dynamics on the probe, although it already differs from the pure equilibrium case: the friction on the probe is reduced by the resetting and the fluctuation--dissipation relation is broken. That motivates the introduction of an effective temperature in that case. In a second regime the influence of the nonequilibrium becomes more severe: the noise felt by the probes becomes nonGaussian, and can even exhibit power-law distributed jumps (third regime). That occurs when the bath would be unstable in the absence of resetting: the resetting stabilizes the bath particles against an inverted harmonic well, a mechanism that produces heavy tails in the position distribution of the bath particles. \\

We start in the next section with the model, and point out what quantities in the probe-bath system are relevant for the induced motion. The linearity of the model replaces the more general regime of linear response around the bath nonequilibrium steady condition, as was outlined first in \cite{jsp}.   
In Section \ref{mre} we give the main results in the scaling regime of infinite baths. The detailed calculations follow in Section \ref{Sec:SingleParticleBathCase} with the rescaling in Section \ref{Sec:ManyParticleBath}.

\section{Model and questions}
We start with the set-up which will provide the logic of the arguments and will point to the calculations that need to be performed.

\subsection{Coupled dynamics}
For simplicity we use one-dimensional notation.\\
A point probe at position $q_t\in \JR$ interacts with a large number $N$ of bath particles at positions $x^i_t\in \JR$
following the potential,
\bea
\text{U}(x,q) = \sum_{i} U(x^i , q) \quad , \quad U(x,q) = \frac{a_1}{2} q^2 + \frac{a_2}{2} x^2 + \frac{a_3}{2} (x-q)^2 \, .\label{ene}
\eea
All interactions are linear and the probe is confined by talking $a_1 + a_3 >0$. The parameters $a_2$ and $a_3$ are not necessarily both positive, and we will be interested in the case where $b=a_2+a_3 < 0$.
 The force between probe and bath is via the average position of bath particles $ \sum_{i=1}^N x^i$ so that we may think of $q$ as being trapped near the center of mass of the bath when $a_3>0$ is large.\\

The bath dynamics is described by an overdamped diffusion with random resetting, \cite{ss,ma2,ma1,ma3,ma4,rold}.  For the latter we pick a rate $r\geq 0$ to select a sequence of random times $t_k^i$, independently for each bath particle, with waiting times $t_{k+1}^i - t_k^i$ exponentially distributed with rate $r$. At these times for that particle we reset its position at a fixed position $A$.   In other words, at each of these times, the particle then say at $x^i$ moves/jumps instantaneously to $x=A$. The equation of motion is therefore, for $i=1,\ldots, N$,
\begin{equation}\label{odm}
 \gamma \dot{x}^i_t = - a_2\, x^i_t - a_3\,(x^i_t-q_t)+ \sqrt{2 \gamma T}\, \xi^i_t + \sum_{k}  \gamma \delta(t-t_k^i) (A - x_t^i) \, ,
\end{equation}
with damping coefficient $\gamma$ and environment temperature $T$. 
There is no direct mutual interaction between the bath particles but there is with the probe at position $q_t$ at time $t$.\\

 The probe dynamics is Newtonian,
\begin{equation}\label{pd}
M \ddot{q}_t - g_t(\dot{q}_t,q_t) = - a_1 q_t - a_3 \sum_{i=1}^{N} (q_t- x_t^i ) \, ,
\end{equation}
where $g_t$ is an additional arbitrary time-dependent force, possibly also random but which plays no further role for the present paper. The objective is to ``integrate out'' the $N$ bath particles, where we will decide later about the time ($t$) and energy scale ($M$) to be considered. \\

\begin{figure}
\centerline{\includegraphics[width=12cm]{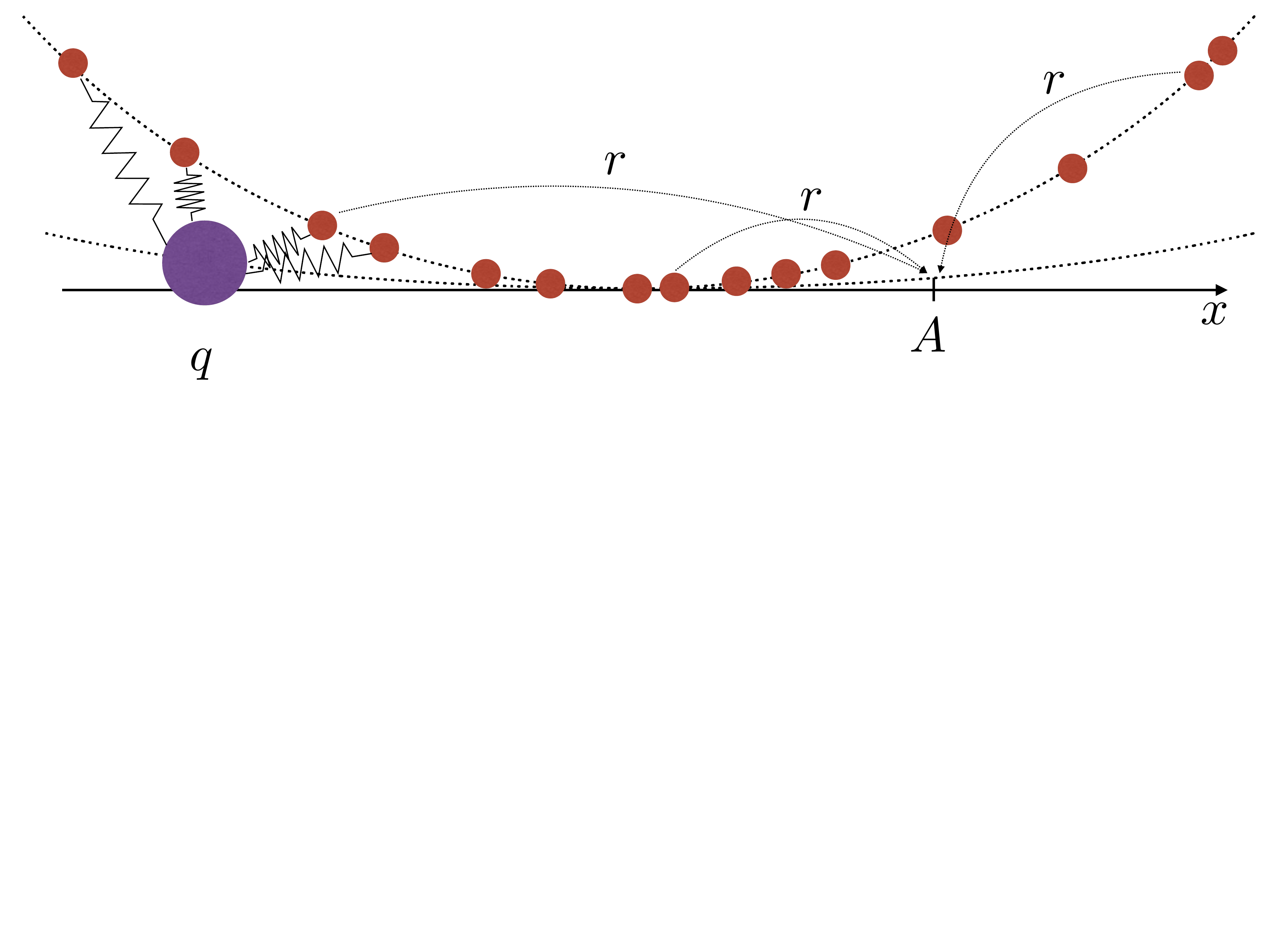}} 
\caption{A probe (purple) at position $q$ interacts linearly with a large number of bath particles (red) and an external parabolic well (that can be replaced by an arbitrary potential). The bath particles feel a parabolic well (eventually repulsive) centered around the origin $x=0$,  are in contact with an external (equilibrium) bath at temperature $T$ and are driven out-of-equilibrium by a resetting process: with rate $r$ the position of each bath particles is independently reset to the position $x=A$.}
\label{fig:Main}
\end{figure}

The equations \eqref{odm}--\eqref{pd} specify the coupled model dynamics.   See Fig.~\ref{fig:Main} for a cartoon of the coupled system.  The initial condition is far in the past, say at time $-\infty$, so that we only care here for the stationary nonequilibrium condition. When the probe is released and interacts with the bath the particles there back-react to its motion providing the statistical forces that determine the induced law of the probe's stochastic motion. 

\subsection{The induced motion}
Let $\langle x_t^i | \{q_{t'}\}^t_{-\infty}\rangle$ denote the average bath-particle position given the probe's history from arbitrarily far in the past. For obtaining the probe's induced motion from \eqref{pd} we use the strategy of \cite{jsp} and we start by the decomposition,
\begin{equation}\label{deo}
x_t^i = \langle x_t^i | \{q_{t'}\}^t_{-\infty}\rangle + \eta_t^i, \quad \eta_t^i:= x_t^i - \langle x_t^i | \{q_{t'}\}^t_{-\infty}\rangle  \, ,
\end{equation} 
formally leading to a reduced probe dynamics (assuming $g_t\equiv0$ in \eqref{pd}),
\bea\label{deo2}
M \ddot{q_t} = -(a_1 +N a_3) q_t  + a_3  N \langle x_t^i | \{q_{t'}\}^t_{-\infty}\rangle + a_3 \sum_{i=1}^N \eta_t^i \, .
\eea
The physics behind that decomposition is that $\langle x_t^i | \{q_{t'}\}^t_{-\infty}\rangle$ corresponds to the systematic force exerted by the out-of-equilibrium bath on the probe, containing notably the thermodynamic force and the friction, while $\eta_t^i$ is the fluctuating force.
The noise $\eta_t^i$, with mean zero, is the result of a last resetting of the $i-$th particle at random time $s(t)<t$ after which it evolves under thermal noise $\xi_{t'}, s(t)\leq t'\leq t$, following equation
\begin{equation}\label{odma}
 \gamma \dot{x}^i_t = - a_2\, x^i_t - a_3\,(x^i_t-q_t)+ \sqrt{2 \gamma T}\, \xi^i_t
  \end{equation}
(equation \eqref{odm} without last sum).
Under some scaling limit the sum of noises in \eqref{deo2} will converge to the noise on the probe.  That noise is Gaussian when $a_2 + a_3 +\gamma\,r/2 >0$, and the covariance will be calculated explicitly in Section \ref{var}.  If not (a case that only exists in the presence of the nonequilibrium, $r>0$), the second moment in the stationary distribution of the bath particles diverges and the noise is another stable distribution (requiring another scaling procedure). In the most severe case $a_2 + a_3 +\gamma\,r \leq  0$ the averaged value $\langle x_t^i | \{q_{t'}\}^t_{-\infty}\rangle$ also diverges and we have to abandon the decomposition \eqref{deo}, the noise on the probe then exhibits power-law distributed jumps. Details will be given in the next section. \\

For estimating $\langle x_t^i | \{q_{t'}\}^t_{-\infty}\rangle$ (when it exists), we use that for $r>0$ each bath-particle has been reset at $A$ with probability one, so that
\begin{equation}\label{est}
\left\langle x_t^i | \{q_{t'}\}^t_{-\infty}\right\rangle = \int_{-\infty}^t \id s \,r\,e^{-(t-s)r}\,\langle x_t^i | \{q_{t'}\}^t_{s}\rangle^0 \, ,
\end{equation}
where the time $s=s(t)$ in the integration stands for the last resetting time of the $i-$th particle and the last expectation $\langle \cdot\rangle^0$ in the integrand is for the undriven dynamics
\eqref{odma} started at time $s$ with $x_s^i=A$. We thus need to know the right-hand side of \eqref{est} which is possible to compute exactly in this linear dynamics.\\

The main interest is to have an exactly solvable model of a probe motion in contact with a nonequilibrium bath.  The typical question that arises here is to see what are the possible differences from the equilibrium case, in terms of friction, noise and stability.

\section{Main results}\label{mre}

We present here the main findings; the remaining sections give the detailed derivations.

\subsection{Stationary bath distribution}

Assuming that the probe is fixed at some position $q_t=q$ in \eqref{odm}, the position of the bath particle reaches at large times the stationary distribution\footnote{The (non-stationary) distribution of $x_t$ conditioned on the probe history is obtained by replacing $q \frac{\lambda}{b} (1 - e^{-\frac{\tau b}{\gamma}}) \to  \frac{\lambda}{\gamma} \int_{0}^\tau  dt' q_{t-t'} e^{-\frac{t' b}{\gamma}}$ in the exponential in \eqref{Eq:ProbabilityDistribution}.},
\bea \label{Eq:ProbabilityDistribution}
p_\text{stat}(x|q) = \int_{0}^{\infty} r \,\id \tau \,e^{-r\tau}  \frac{1}{\sqrt{2\pi\, T \,(1- e^{- 2\frac{\tau b}{\gamma}})/b}} \exp\{- \frac{(x-A e^{- \frac{\tau b}{\gamma}} - q \frac{\lambda}{b} (1 - e^{-\frac{\tau b}{\gamma}}))^2}{ 2T (1- e^{- 2\frac{b\tau}{\gamma}})/b}\} \, ,
\eea
where we have introduced $b=a_2+a_3$ and $\lambda=a_3$. That defines a probability distribution whenever the resetting rate $r>0$, even if $b \leq 0$. In that last case the bath particle is stabilized by the resetting. Indeed, irrespective of the sign of $b$, the argument of the last exponential in \eqref{Eq:ProbabilityDistribution} converges to a constant for large $\tau$, and the convergence is thus ensured by the first exponential $e^{-r\tau}$. We note that the distribution \eqref{Eq:ProbabilityDistribution} was already obtained in \cite{pal} where stochastic resetting in various potential landscapes was considered. For $x\rightarrow A$, there is always a cusp.  The resetting induces a  jump in the first derivative: the coefficients $c_{\pm} = \lim_{x \to A^\pm} \partial_x p_\text{stat}(x|q ,A)$ satisfy $c_+ - c_- = -2 \times \frac{r \gamma}{2T}$, independently of all the other parameters but with in general $c_+ \neq - c_-$. For $q=0=A$, $p_\text{stat}(x=0|q = 0 ,A=0) - p_\text{stat}(x|q = 0 ,A=0) \simeq \frac{r \gamma}{2T} |x| $. That follows from studying the stationary Fokker-Planck equation satisfied by $p_\text{stat}$.\\

\begin{figure}
\centerline{\includegraphics[width=16cm]{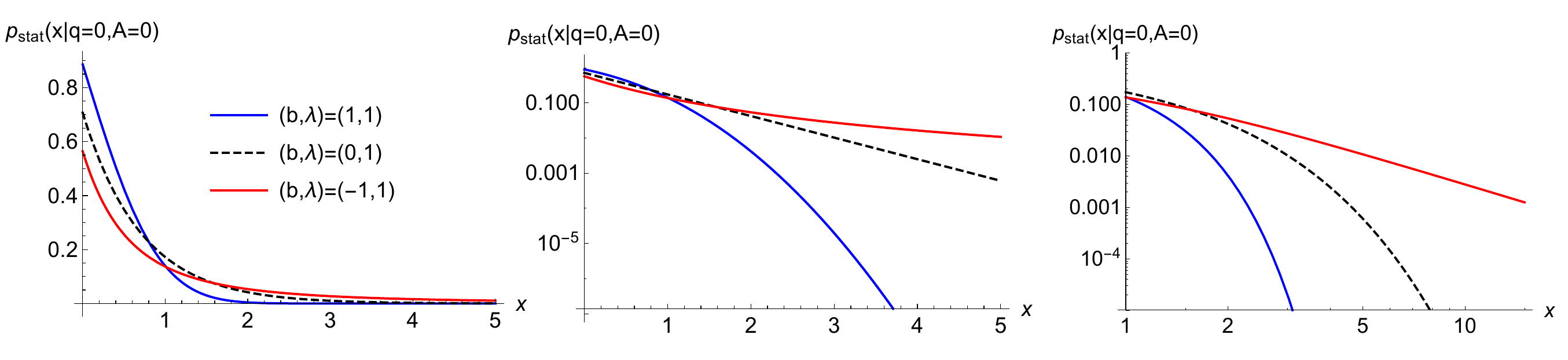}} 
\caption{Plot of the stationary distribution for the bath particles when both the probes and the resetting are at the origin $q=A=0$ in linear scales (left), logarithmic scale (middle) and log-log scale (right), for all parameters set to $1$ and $b=1$ (blue line), $b=0$ (black dashed line) and $b=-1$ (red line). For $b>0$ the large $x$ behavior is Gaussian-like, while for $b<0$ the distribution exhibits a fat-tail. In the case $b=0$ the decay is simply exponential.}
\label{fig:plot123}
\end{figure}

The bulk and asymptotic properties of that nonequilibrium distribution however strongly depend on the sign of $b$.  Remember that $b=a_2 + a_3$ and $bx^2/2$ in \eqref{ene} gives the harmonic (anti-)well for the bath particles near the origin.\\
   Consider first the case $q=A=0$ and $b \neq 0$. If $b>0$, then one gets a Gaussian decay $p_\text{stat}(x|q = 0 ,A=0) \sim  e^{-b\frac{x^2}{2T}}$ (up to subdominant terms) which is the same as in equilibrium ($r=0$).\\
    If $b <0$ (where the bath would be unstable under \eqref{odma}), then,
   \bea \label{Eq:FatTail}
   p_\text{stat}(x|q = 0 ,A=0)  \sim_{|x| \to \infty} G \frac{\gamma r/|b|}{(2T/|b|)^{\frac{\gamma r}{2b}}}\,\,\frac{1}{|x|^{1 -\gamma r/b}}  \, ,
   \eea
   with $G = \frac{1}{2 \sqrt{\pi}} \int_{0}^{+\infty}  \id y y^{\frac{\gamma r/b-3}{2}} e^{-1/y}  = \frac{\Gamma(\frac{3-\gamma r/b}{2})}{2\sqrt{\pi}}   $. The distribution thus develops a fat tail when $b <0$, as the result of a competition between the distance $\Delta x \sim \sqrt{\frac{T}{|b|}} e^{|b|\tau/\gamma}$ traveled by the bath particle between two resetting events at time $t$ and $t+\tau$, and the probability $\sim e^{-r\tau}$ to observe such a resetting. Note that the first, respectively the second moment of the stationary distribution thus ceases to exist when $b$ is too negative, more precisely when $r\gamma+b \leq 0$, respectively  $r\gamma+2b \leq 0$.  Finally the (physically relevant) intermediary case is obtained taking $b \to 0^+$. In that case we obtain
   \bea
   p_\text{stat}(x|q = 0 ,A=0)  =\frac{\gamma r }{4T\sqrt{\pi}}  \int_{0}^{\infty} \frac{\id y}{\sqrt{y}}\, e^{-\frac{x^2}{y}  - \frac{r \gamma}{4T}y} = \sqrt{\frac{r\gamma}{4T}}\, e^{-\sqrt{\frac{\gamma r}{T}} |x| } \, ,
   \eea
   the known result for the stationary distribution of a free particle under resetting \cite{ma2}. The distribution $p_\text{stat}(x|q = 0 ,A=0)$ is plotted in Fig.~\ref{fig:plot123} for various choices of parameters.\\

\begin{figure}
\centerline{\includegraphics[width=5.5cm]{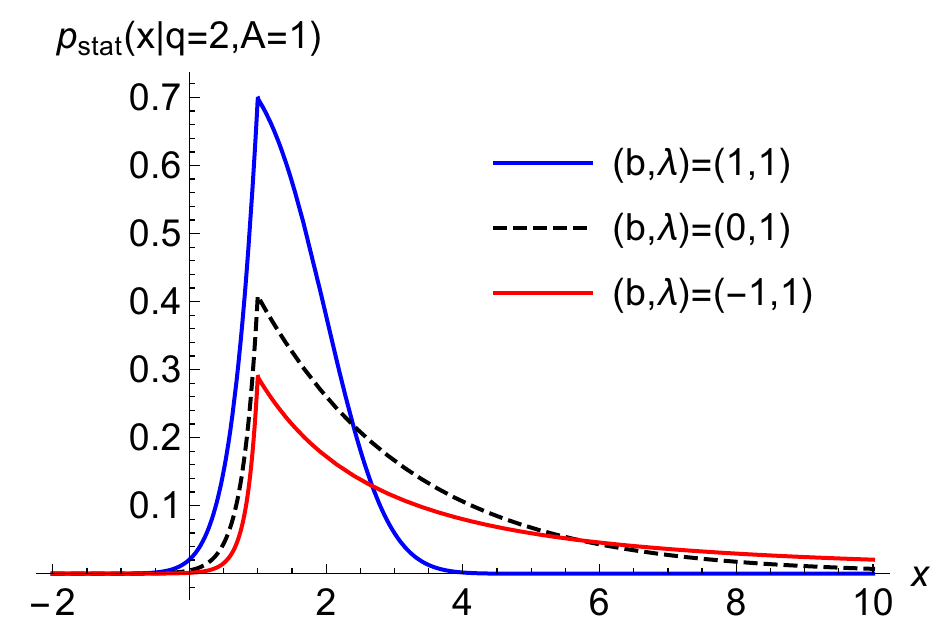}  \quad \includegraphics[width=5.5cm]{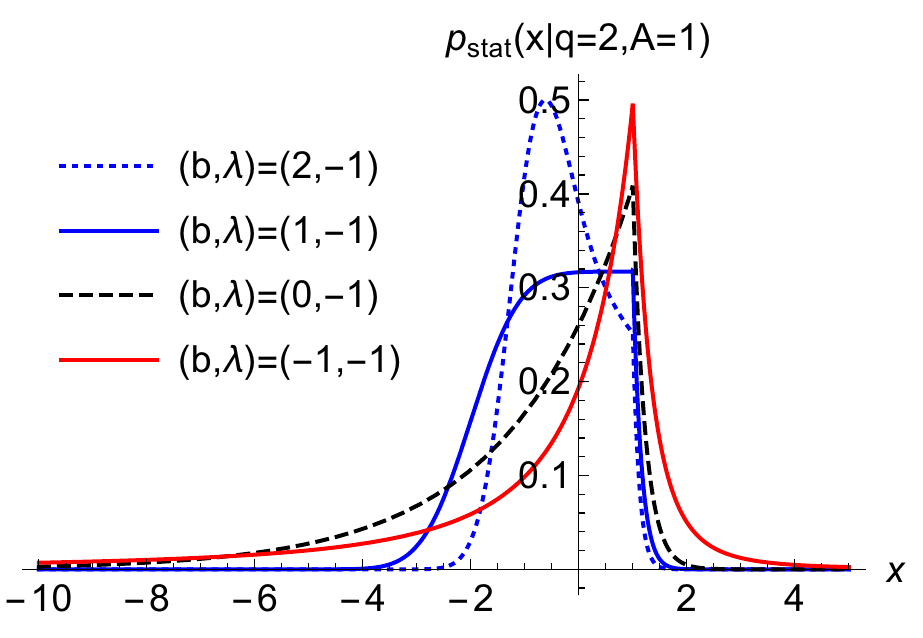}}  
\caption{Plot of the stationary distribution for the bath particles for $q=2$ and $A=1$. All parameters of the models are set to unity except $\lambda=1$ on the left, and $\lambda=-1$ (repulsive probe-bath interaction) on the right; $b=1$ (blue line), $b=0$ (lack-dashed line) and $b=-1$ (red line), with also $b=2$ on the right (blue dotted line).}
\label{fig:plot45}
\end{figure}

     The general case $q\neq 0\neq A$ adds various complications and the distribution is not symmetric anymore; see Fig.~\ref{fig:plot45} for some examples.  The large $x$ behavior remains qualitatively similar but the tails are now asymmetric ({\it i.e.,} the prefactors depend on the direction). For example, in the case $b<0$ we have in general 
   \bea \label{Eq:FatTail2}
   p_\text{stat}(x|q  ,A)  \sim_{ x  \to \pm \infty} G_\pm \frac{\gamma r/|b|}{(2T/|b|)^{\frac{\gamma r}{2b}}} \,\,\frac{1}{|x|^{1 -\gamma r/b}}  \, ,
   \eea
   with $G_{\pm} = \frac{1}{2 \sqrt{\pi}} \int_{0}^{+\infty}  dy y^{\frac{\gamma r/b-3}{2}} e^{- \frac{1}{y}(1 \mp (A - \frac{q\lambda}{b}) \sqrt{\frac{-b y}{2T}})^2} $. If $0<q \lambda -Ab$ (which is the force felt by the bath particle just after resetting), then $G_+ >G_-$ as it is more likely for the bath particle to quickly diverge (under the effect of the repulsive parabolic well) towards the region $x>0$.

\subsection{Exact reduced probe dynamics in the large $N$ limit}

\subsubsection{Gaussian case}
We first discuss the case where $r\gamma + 2b >0$, $b = a_2 + a_3$.  We rescale
\bea \label{Eq:ScalingLimit}
a_3 = a_3'/N \quad , \quad a_2 = a_2'/N \quad , \quad t = N t'  \quad , \quad  r =  r'/N  \quad , \quad M = N^2 M' \, ,
\eea
for which $r'\,t' = r\,t$: the rate at which the bath particles are resetted is kept constant. We obtain the reduced probe dynamics exactly in that scaling limit. Our result is that $Q_{t'} := \lim_{N \to \infty} q_{N t'}$ follows the stochastic equation of motion (assuming $g_t\equiv0$ in \eqref{pd})
\bea\label{redd}
M' \ddot{Q} + \int_0^{+\infty}\id{\tau}\, K_{r'}'(\tau) \dot{Q}_{t-\tau} = F_\text{tot}(Q_t)  + \zeta_t \, ,
\eea
where for each $r\geq 0$ ($r=0$ being the equilibrium case, also included in this result),
\begin{itemize}
    \item
The total force acting on the particle is 
\bea  \label{Eq:IntroFtot}
&& F_\text{tot}(Q) = - a_1 Q  -\frac{a_2' a_3'}{a_2' +a_3'} Q  - \frac{a_3'^2}{a_2'+a_3'}  \frac{\gamma r'}{\gamma r' + a_2'+a_3'}  \left( Q -   \frac{a_2'+a_3' }{a_3' } A \right)   \, . 
\eea
The first and second terms on the right-hand side can be identified as the systematic (thermodynamic) force exerted by the bath particles on the probe, equal to $-\partial_Q {\cal F}[T,Q]$ with ${\cal F}[T,Q]$ the equilibrium free-energy (per particle) of the bath with the probe fixed at position $q$: ${\cal F}[T,q]= \frac{a_3\,a_2}{2 (a_2+a_3)}  q^2 - \frac{T}{2} \log( \frac{2 \pi T}{a_2+a_3})$. The third term is only present in the out-of-equilibrium $r>0$ case and cannot simply be related to a free-energy function; see \cite{prl}. If we had $a_2<0$ with still $ a_2 + a_3 >0$  so that for small $a_1>0$ the origin would be an unstable fixed point for the probe dynamics $M' \ddot{Q}  = F_\text{tot}(Q_t)$ in the case $r=0$ (equilibrium), a resetting process at the origin (third term with $A=0$) could stabilize it. If $A \neq 0$ the equilibrium position $Q^*$ defined by $F_\text{tot}(Q^*) = 0$ is shifted from the origin.
    \item
The friction kernel is
\bea\label{frk}
K_{r'}'(t) =  \frac{a_3'^2}{a_2'+a_3' + \gamma r'}  e^{- (r' + \frac{a_2'+a_3'}{\gamma}) t},\qquad t\geq 0
\eea
with the resetting parameter $r'$ lowering the overall amplitude of the friction kernel and the correlation time.  That is an example of out-of-equilibrium shear thinning, as the probe moves more easily through the medium when the resetting rate $r$ grows.
    \item
    The noise $\zeta_t$ is Gaussian with mean zero and covariance
    \bea \label{Eq:Intro:NoiseCorrelations}
    \langle \zeta_t\, \zeta_{t'} \rangle = T\, \frac{a_3'^2}{a_2' +a_3' +\gamma r'/2}  e^{- (r' + \frac{a_2'+a_3'}{\gamma})|t-t'|} \label{rdf} \, .
\eea
There is the equilibrium decorrelation time $\gamma/(a_2' + a_3')$ to which is added the time-scale of the resetting, where the noise becomes more white for larger resetting rate $r$. The amplitude of the noise suggests to introduce an effective temperature as we show in Section \eqref{eft}.
\end{itemize}

\subsubsection{NonGaussian cases}\label{ngc}
When $a_2 + a_3 =b<0$ and $ -b< \gamma r < -2b $, the fat tail \eqref{Eq:FatTail} in the distribution of bath particles renders the second moment of $x$ divergent while the first moment remains finite\footnote{ The transition point $\gamma r =-2b$ requires some special care and is not considered here.}. The sum of noises $\zeta_t = \lambda \sum_{i=1}^N \eta_t^i$ can still be rescaled however to converge to a nonGaussian noise whose one-time distribution is a generalized stable distribution (see e.g. \cite{GCLT}). We now take a scaling limit with the fat-tail exponent 
\bea
\alpha = - \frac{\gamma r}{b}  
\eea 
of the one-time distribution of $\eta_t^i$ fixed. We choose $a_2 \sim  \frac{a_2'}{N^{2/\alpha}}$, $a_3=\lambda \sim  \frac{a_3'}{N^{2/\alpha}}$, $r \sim \frac{r'}{N^{2/\alpha}}$, $t \sim t' N^{2/\alpha} {}$ and $M \sim M' N^{4/\alpha}$. Under that rescaling the systematic part of the reduced dynamics (force and friction) converges to $0$ and we obtain in the limit  (assuming $g_t\equiv 0$ in \eqref{pd}) the equation,
\bea\label{14}
M' \ddot{Q}_t = - a_1 Q_t + \zeta_t \, ,
\eea
with $\zeta_t$ a noise such that its one time distribution is a generalized $\alpha$-stable distribution $S(\alpha , 0 , c , 0)$ with scale factor
\bea \label{Eq:StableDistc}
c = \left(  \frac{ \sqrt{\pi} \Gamma((3+\alpha)/2)}{2 \sin(\frac{\pi \alpha}{2}) \Gamma(\alpha))}  \right)^{1/\alpha}  \sqrt{ \frac{2 T (a_3')^2}{|a_2'+a_3'|}  } \, .
\eea
That means that the characteristic function of the noise is $\langle e^{i \omega \zeta_t } \rangle =e^{ - c^{\alpha} |\omega|^{\alpha}} $. The result \eqref{14} remains true even when $\gamma r \leq -b$ and the decomposition \eqref{deo} does not make sense (because $\langle x^i_t \rangle = \infty$).  Then, the noise is given by $\zeta_t = \lim_N \frac{a_3'}{N^{2/\alpha}} \sum_{i=1}^N x_t^i$ and still has the law $S(\alpha , 0 , c , 0)$. In general there is no known explicit expression for the probability density $p(\zeta)$ of the distribution $S(\alpha , 0 , c , 0)$ beyond its asymptotic decay $p(\zeta) \sim_{|\zeta| \to \infty} \frac{c^{\alpha} \sin(\frac{\pi \alpha}{2} )\Gamma(1+\alpha)}{\pi |\zeta|^{1+\alpha}}$. A notable exception is the case $\alpha=1$ where one obtains the Cauchy distribution with $p(\zeta) = \frac{1}{\pi c(1+\zeta^2/c^2)}$.

\begin{figure}
\centerline{\includegraphics[width=13cm]{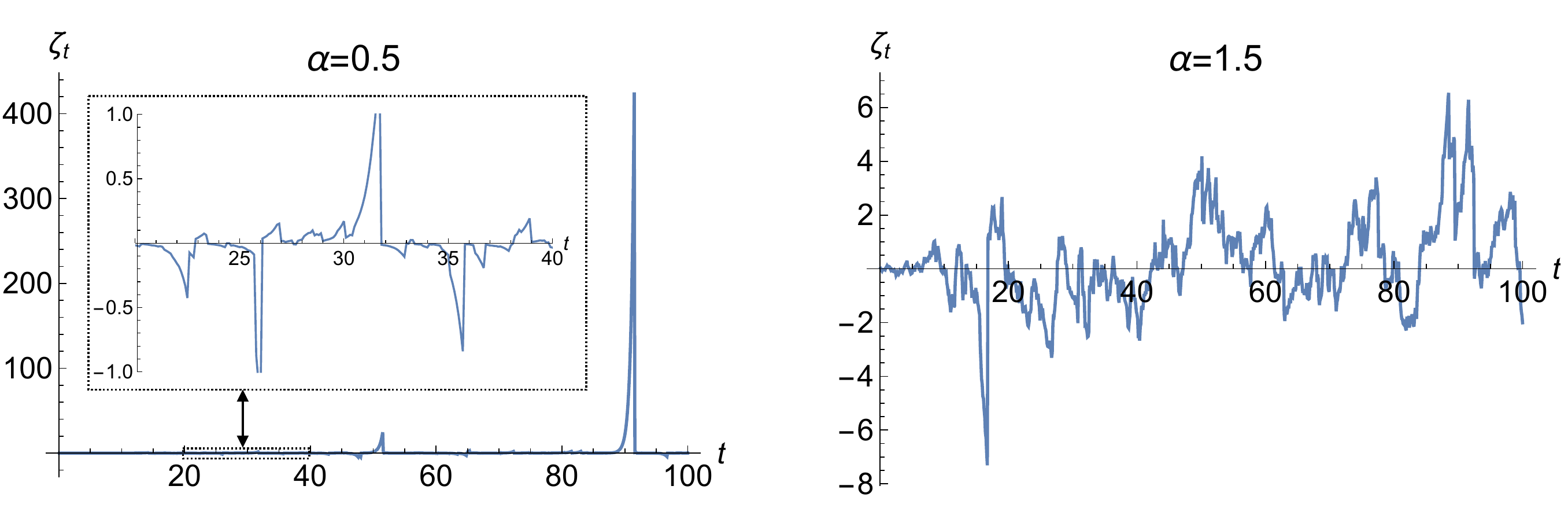}}  
\caption{Simulation of the (rescaled) noise felt by the probe with here $A=q=0$ fixed, all the (rescaled) parameters set to unity except $a_2+a_3$ that is fixed to ensure the fat-tail exponent $\alpha = 0.5$ (left) or $\alpha = 1.5$ (right). Here we use $N=10^6$ particles. In both cases the one-time distribution of the noise is non-Gaussian with a fat-tail $p(\zeta)\sim \zeta^{-1-\alpha}$. In the case $\alpha >1$ the noise is continuous. In the case $\alpha<1$ it exhibits jumps that are power-law distributed with the same tail exponent $\alpha$.}
\label{fig:plot6}
\end{figure}

The characterization of the full process (in particular its time-correlations) is more difficult. Still, we note that its nature strongly changes at $\alpha = 1$  $(\gamma r = -b)$. When $\alpha < 1$ (i.e., $\gamma r < -b$) the process $\zeta_t$ exhibits power-law (with the same tail exponent $\alpha$) distributed jumps. Indeed in that regime a single particle, e.g. the one that is the furthest away from the origin, can contribute to a finite fraction of the noise $\lambda \sum_{i=1}^N x_t^i = \zeta_t$: $\zeta_t$  and $\lambda \text{max}_i | x_t^i|$ are of the same order (i.e. $O(1)$ with the rescaling). That is the `condensation' phenomenon for the sum of power-law distributed random variables; see e.g. \cite{Condensation,Condensation2}. Then, at resettting the position of that particle  at time $t_r$ (which occurs at least with rate $r$), the noise $\zeta_t$ changes by a finite amount $ \zeta_{t_r^+}^i-\zeta_{t_r^-}^i =   A -\lambda \text{max}_i | x_t^i| = O(1)$, a jump which is power-law distributed with the same exponent $\alpha$ (the distribution of the maximum of power-law distributed random variables is a Fr\'echet distribution displaying the same power law tail). In the case $\alpha >1$ on the other hand, the noise $\zeta_t$ is continuous at large $N$, as in the Gaussian case. Indeed in that case each noise $\eta^i_t$ contributes to a vanishing fraction (in the large $N$ limit, for example the contribution of the maximum is now $O(N^{\frac{1-\alpha}{\alpha}})$) of the noise $\zeta_t$ and the jumps in the individual noises $\eta^i_t$ (still stemming from the resetting) are washed out in the total noise felt by the probe. A plot of some typical realizations of the noise in the case $\alpha<1$ and $\alpha>1$ is given in Fig.~\ref{fig:plot6}.

\subsection{Effective temperature and relaxation in the Gaussian case}\label{eft}

We go back to the case where $2(a_2+a_3) + \gamma r>0$.\\
A salient signature of the nonequilibrium nature of the induced probe dynamics is that the fluctuation--dissipation theorem is broken by the resetting. Taking the ratio of the noise correlations and of the friction kernel leads to the definition of an effective temperature
\bea \label{Eq:IntroTeff}
T_\text{eff} && := \frac{\langle \zeta_\alpha(t) \zeta_\alpha(t-\tau) \rangle}{K_{r'}'(\tau) }  = T \frac{a_2 +a_3 +\gamma r}{a_2 +a_3 +\gamma r/2}  > T   \, ,
\eea
that is larger than the temperature of the external equilibrium bath whenever $r>0$. We can also define the effective potential felt by the probe
\bea \label{Eq:Intro:Veff}
V_\text{eff}(Q) := \frac{1}{2}\left(  a_1 +\frac{a_2' a_3'}{a_2' +a_3'} \right) Q^2  + \frac{1}{2} \frac{a_3'^2}{a_2'+a_3'}  \frac{\gamma r'}{\gamma r' + a_2'+a_3'}  \left( Q -   \frac{a_2'+a_3' }{a_3' } A \right)^2  \, ,
\eea
for which $F_\text{tot}(Q) = -\partial_Q V_\text{eff}(Q)$.  As a consequence the probe reaches the stationary distribution $p(Q)  \sim e^{-\frac{V_\text{eff}(Q)}{T_\text{eff}} }$.  Nothing fundamentally distinguishes the probe dynamics from a dynamics that is obtained in the absence of resetting, except if one also has access to {\it e.g.} the real temperature.\\

We can also ask what is the most efficient (more precisely, the fastest) way to reach a given stationary distribution for the probe if all parameters are fixed except the resetting rate $r$ and the global confining potential parameters $a_1$ and $a_2$. We thus take $A=0$ and keep constant $\gamma$, $T$ and the length scale $\ell$ defined by
\bea \label{Eq:ConstraintTherm}
\ell^2 = \frac{T_\text{eff}(r,a_i)}{\kappa_\text{eff}(r,a_i)} \, ,
\eea
with $T_\text{eff}$ given in \eqref{Eq:IntroTeff} and $\kappa_\text{eff}=a_1 +\frac{a_2' a_3'}{a_2' +a_3'} + \frac{a_3'^2}{a_2'+a_3'}  \frac{\gamma r'}{\gamma r' + a_2'+a_3'}$ the stiffness of the effective confining potential \eqref{Eq:Intro:Veff}. $\ell$ controls the width of the stationary distribution of the probe $p(Q) \sim e^{-Q^2/(2\ell^2)}$. To evaluate the relaxation time of the probe we also need the effective friction coeffitient defined by (dropping the primes everywhere now),
\bea
\gamma(r,a_i) = \int_0^{+\infty}\id\tau\,K_{r}(\tau) =\gamma  \frac{a_3^2}{(a_2+a_3+\gamma r)^2} \, .
\eea
In a Markovian approximation of the reduced dynamics \eqref{redd} the relaxation time of the probe is then given by
\bea
\tau(r,a_i) = \frac{\gamma(r,a_i)}{\kappa(r,a_i)} = \frac{\gamma \ell^2}{T}  \frac{a_3^2 (a_2+a_3+\gamma r /2)}{(a_2+a_3+\gamma r)^3} \, ,
\eea
and one should remember that there is a relation linking the $a_1$, $a_2$ and $r$ in this setting \eqref{Eq:ConstraintTherm}. Note that at equilibrium $r=0$, $\tau = \frac{\gamma \ell^2}{T} a_3^2$ is independent of the stiffness of the external parabolic wells $a_1$ and $a_2$. Out--of--equilibrium one gets a dependence on the three `confining' parameters $(r,a_1,a_2)$ and it is now possible to tune the relaxation time of the probe. In particular, keeping $a_2$ constant and tuning $a_1$ as a function of $r$ so as to keep the constraint \eqref{Eq:ConstraintTherm} it is clear that one gets a relaxation time that strictly decays with $r$ and for $r \to \infty$,  $\tau(r,a_i)  \simeq \frac{\gamma \ell^2}{2T} \frac{a_3^2}{(\gamma r)^2} $. Keeping $a_1$ constant and tuning $a_2$ to satisfy \eqref{Eq:ConstraintTherm} leads to a more complicated dependence of $\tau(r,a_i)$ on $r$ but we always find that the relaxation time monotonically decreases with $r$.

\section{Single particle bath case} \label{Sec:SingleParticleBathCase}

The computations that lead to the previously announced results all depend on calculations for a single bath-particle interacting with the probe on which we focus in this section.

\subsection{Solution for the bath particle between resetting times}

For $N=1$ (one bath-particle with position $x_t$) we have the coupled equations of motion \eqref{odm}--\eqref{pd} that we rewrite here for clarity (with $g_t\equiv 0$), in terms of the parameters $b=a_2+a_3$, $\kappa=a_1+a_3$ and $\lambda=a_3 $,
\bea \label{Eq:1ParticleAndProbeLinearDynamics}
&& M \ddot{q}_t = - \kappa q_t + \lambda x_t \nonumber \\
&& \gamma \dot{x}_t =  - b x_t + \lambda q_t + \sqrt{2 \gamma T} \xi_t + \sum_{k}  \gamma \delta(t-t_k) (A - x_t)  \label{sbp} \ .
\eea
We remember that $\xi_t$ is a standard white noise $\langle \xi_{t} \xi_{t'} \rangle = \delta(t-t')$, and the $t_k$ are positive random variables which are just the times at which an exponential clock with rate $r>0$ rings: $t_0$ has a pdf $p(t_0) = r e^{- r t_0}$ and the waiting times $\Delta t_i =t_i -t_{i-1}$, $i \geq 1$ are also iid with an exponential distribution $p(\Delta t_i) = r e^{- r \Delta t_i}$.  We take the resetting to the position $A$.

When the last resetting happened at time $s<t$, the solution of \eqref{sbp} is, given the probe position $q_t$
\bea \label{Eq:Sol:OrnsteinUhlenbeck}
 x_t  =  A e^{-\frac{t-t'}{\gamma_b}}+ \int_{s}^{t} (\frac{q_{t'}}{\gamma_{\lambda}}  + \sqrt{D}\xi_{t'} ) e^{-\frac{t-t'}{\gamma_b}} dt' \, ,
\eea
where we introduced the time scales $\gamma_{b} = \gamma/b$ and $\gamma_{\lambda} = \gamma/\lambda$ and the diffusion coefficient $D = 2T/\gamma$. 
Hence $(x_{t'})_{t' \in [s,t]}$ is a Gaussian stochastic process fully characterized by its first two cumulants
\bea \label{Eq:DefPhi1Phi2}
&& \phi_1(s,A;t)  := \langle x_t  |\{q_{t'}\}_s^t \rangle = A e^{-\frac{t-t'}{\gamma_b}}+ \int_{s}^{t} \frac{q_{t'}}{\gamma_{\lambda}}e^{-\frac{t-t'}{\gamma_b}} dt' \, ,  \\
&& \phi_2(s;t_1,t_2) :=  \langle x_{t_1} x_{t_2} \rangle_c =_{t_1 \leq t_2}  D \int_{s}^{t_1} e^{-  \frac{t_1+t_2 - 2t'}{\gamma_b}}  dt'  = \frac{D \gamma_b}{2} \left( e^{- \frac{|t_2-t_1|}{\gamma_b}}  - e^{- \frac{t_1+t_2 - 2 s}{\gamma_b}} \right) \, , \nonumber
\eea
all assuming that the last resetting was at time $s<t_1,t_2,t$.  The $\phi_1$ and $\phi_2$ will be the building blocks of our computations, and the notation $\{q_{t'}\}_s^t$ emphasizes the dependence on the probe trajectory $q_{t'}$ for $t' \in [s,t]$.

\subsection{Stationary distribution of the bath particle}
Here we supose that $q_t=q$ is fixed. To get the distribution of $x_t$, note that conditioned on the last resetting time having occured at time $s$, this distribution is Gaussian with the moments \eqref{Eq:DefPhi1Phi2}. Integrating over the distribution of $\tau =t-s$ (i.e. the time since the last resetting) we get
\bea
p_\text{stat}(x|q) = \int_{0}^{+\infty} r d\tau e^{-r\tau} \frac{1}{\sqrt{2 \pi \phi_2(t-\tau;t,t)}} e^{- \frac{(x-\phi_1(t-\tau,A;t))^2}{2 \phi_2(t-\tau;t,t)}} \, .
\eea
That implies  \eqref{Eq:ProbabilityDistribution}, the stationary distribution of an Onstein-Uhlenbeck process with resetting as already studied in \cite{pal}.
We next perform the change of variables $y =  2T (1- e^{-2 \tau b/\gamma})/b \geq 0$. We get for \eqref{Eq:ProbabilityDistribution},

\begin{equation}
p_\text{stat}(x|q = 0 ,A=0) = \frac{r\, \gamma}{4T}\,\int_{0}^{C_b}  \id y \,\frac{(1- \frac{b y}{2T})^{-1+ \frac{\gamma r}{2b}}}{ \sqrt{\pi y}}  \exp\{ - \frac{\left(x- \frac{q \lambda}{b} -(A-\frac{q \lambda}{b})\sqrt{1-\frac{b y}{2T}} \right)^2}{y} \} \, ,
\end{equation}
with $C_b = 2T/b$ for $b>0$ and $C_b= +\infty$ for $b <0$. In both cases the integral can be expressed in terms of hypergeometric functions.  Secondly, in the large $x$ limit the integral is dominated by large $y$. If $b>0$, then $y$ is bounded from above and one gets a Gaussian decay.  If $b <0$, then, from rescaling $y \to x^2 y$ we get \eqref{Eq:FatTail} and \eqref{Eq:FatTail2}.

To get the non-analyticity of the stationary distribution around $x=A$, note that the distribution $P(x)=p_\text{stat}(x|A,q)$ must solve the stationary Fokker-Planck equation (this is directly obtained from the dynamics \eqref{Eq:1ParticleAndProbeLinearDynamics})
\bea
0 = -\partial_x \left(\frac{\lambda q - b x}{\gamma} P(x) \right) + \frac{D}{2}\partial_x^2 P(x) - r P(x) + r \delta(x-A) \, .
\eea
And the Dirac distribution must be compensated by a jump in the first derivative at $x=A$: introducing $c_\pm = \lim_{x \to A^\pm} \partial_x P(x) $ we obtain $\frac{D}{2} \partial_x^2 P = \frac{D}{2} (c_+-c_-) \delta(x-A) + \text{more regular terms}$. Hence $c_+-c_- = -\frac{2 r}{D} = -\frac{r \gamma}{T}$.

\subsection{Effective force and friction}

From here we can use \eqref{est} to compute
\bea \label{Eq:FirstMomentExact}
\langle x_t  | \{q_s\}_{-\infty}^t \rangle = \int_0^{+\infty} r \id \tau e^{-r \tau} \phi_1(t-\tau, \{q_{t'}\}_{t-\tau}^t ; t) \, .
\eea
In this simple linear case ``linear response is exact'' and we can compute exactly the first moment in the steady state.
After some calculation we find
\bea \label{Eq:ExactResulKernel}
&& \lambda \langle x_t  | \{q_s\}_{-\infty}^t \rangle  =  F(q_t) - \int_{0}^{\infty} K(\tau) \dot{q}_{t-\tau} \nn \\
&&  F(q) = \lambda \left( A\, \frac{r}{r_b} + \frac{\gamma_b}{\gamma_{\lambda}} \left( 1 - \frac{r}{r_b} \right)  q \right) \nn \\
&& K(t) =  \lambda \frac{\gamma_b}{\gamma_{\lambda}}   \left( 1 - \frac{r}{r_b} \right) e^{- r_b t}  \ ,
\eea
with $r_b = r + 1/\gamma_b$. This formula is valid when $r_b > 0$, otherwise the systematic force is not defined (this is due to the fat tail in the distribution of $x$, see \eqref{Eq:FatTail}). Note that this formula is also valid for $r =0$ (equilibrium case). We see here that the friction term is decreased by the nonequilibrium driving. There are two mechanisms responsible for this decay (i) a shorter memory time $\gamma_b \to \frac{\gamma_b}{1+r \gamma_b}$; (ii) a smaller overall amplitude $\lambda \frac{\gamma_b}{\gamma_{\lambda}} \to \lambda \frac{\gamma_b}{\gamma_{\lambda}}   \left( 1 - \frac{r}{r_b} \right)$.

\subsection{The one-particle noise}\label{var}

For the joint distribution of $x_{t_1},x_t$ for $t_1 \leq t$ we need to consider both  the time $\tau$ of the last resetting before $t$ and $\tau_1$ the time of the last resetting before $t_1$. In general these are not correlated, except if $\tau \geq t-t_1$, in which case $\tau_1 = t_1 - (t- \tau) = t_1 - t + \tau$. We thus have to distinguish two cases which lead to two terms in the correlation function,
\bea
\langle x_{t_1} x_{t} | \{q_s\}_{-\infty}^t\rangle = && \int_{0}^{t-t_1} r e^{-r\tau} \id\tau\int_{0}^{+\infty} re^{-r\tau_1} d\tau_1\phi_1(t_1-\tau_1,A ; t_1) \phi_1(t-\tau,A ; t) \nn \\ 
&& +  \int_{t-t_1}^{+\infty} r \id\tau e^{-r\tau} \left[  \phi_1(t-\tau,A ; t_1)  \phi_1(t-\tau,A ; t)  +  \phi_2(t-\tau; t_1,t) \right] \, , \nn
\eea 
which leads to
\bea
\langle x_{t_1} x_{t} | \{q_s\}_{-\infty}^t \rangle = && \int_{0}^{t-t_1} re^{-r\tau} \id\tau  \phi_1(t-\tau,A ; t) \langle x_{t_1}  |  \{q_s\}_{-\infty}^{t_1} \rangle \nn \\ 
&& +  \int_{t-t_1}^{+\infty} r \id\tau e^{-r\tau} \left[  \phi_1(t-\tau,A ; t_1)  \phi_1(t-\tau,A ; t)  +  \phi_2(t-\tau; t_1,t) \right]  \, . \nn 
\eea 
Note that the first term can be rewritten as
\bea
&& \int_{0}^{t-t_1} re^{-r\tau} \id\tau  \phi_1(t-\tau,A ; t) \langle x_{t_1}  | \{q_s\}_{-\infty}^{t_1} \rangle \nn \\
&& =  \langle x_{t}  |\{q_s\}_{-\infty}^{t_1} \rangle \langle x_{t_1}  |\{q_s\}_{-\infty}^{t_1} \rangle - \int_{t-t_1}^{+\infty}re^{-r\tau} \id\tau  \phi_1(t-\tau,A ; t) \, \langle x_{t_1}  | \{q_s\}_{-\infty}^{t_1} \rangle \, . \nn 
\eea
Hence the connected moment is
\bea
\langle x_{t_1} x_{t} | \{q_s\}_{-\infty}^t \rangle_c = &&  -  \int_{t-t_1}^{+\infty}re^{-r\tau} \id\tau  \phi_1(t-\tau,A ; t) \, \langle x_{t_1}  |\{q_s\}_{-\infty}^{t_1} \rangle \nn \\ 
&& +  \int_{t-t_1}^{+\infty} r \id\tau e^{-r\tau} \left[  \phi_1(t-\tau,A ; t_1)  \phi_1(t-\tau,A ; t)  +  \phi_2(t-\tau; t_1,t) \right]  \, , \nn 
\eea 
and we obtain the decomposition
\bea \label{Eq:DefQ1Q2}
&& \langle x_{t_1} x_{t} | \{q_s\}_{-\infty}^t \rangle_c = Q_1(t-t_1)  + Q_2(t_1,t|\{q_s\}_{-\infty}^{t_1})  \\
&& Q_1(t-t_1) = \int_{t-t_1}^{+\infty} r \id\tau e^{-r\tau}  \phi_2(t-\tau ; t_1,t) = \frac{D \gamma_b}{2} \left( 1- \frac{r}{r_b'} \right) e^{- r_b(t-t_1)}  \nn \\
&& Q_2(t_1,t | \{q_s\}_{-\infty}^{t_1}) = \int_{t-t_1}^{+\infty} r \id\tau e^{-r\tau}  \phi_1(t-\tau,A ; t) \left( \phi_1(t-\tau,A ; t_1) - \langle x_{t_1}  | \{q_{-\infty}^{t_1}\} \rangle  \right)  \, , \nn 
\eea
with $r_b'=r+2/\gamma_b$. That result is only valid in the regime $r_b' >0$. For $r_b' \leq 0$ the second moment does not exist because of the fat tail \eqref{Eq:FatTail} in the distribution of $x_t$. The correlations are generated by two mechanisms: (i) the simple probe-independent correlations that originate from the Gaussian correlations of the free process when there is no resetting time in between $t_1$ and $t$ and that are taken into account in $Q_1(t_1,t)$; (ii) the correlations induced by the correlations between the last resetting times before $t$ and $t_1$ (if $t-\tau < t_1$ then $\tau_1 = t_1- t + \tau$) and that are contained in $Q_2(t_1,t | \{ q_{-\infty}^{t_1} \}) $ and explicitly depend on the probe position before $t_1$.

\smallskip
While an explicit expression for $Q_1$ was already given above, getting an explicit expression for $Q_2$ is more tedious. In the large $N$ scaling limit presented in Section \ref{Sec:ManyParticleBath} only $Q_1$ contributes. At finite $N$, $Q_2$ also contributes. In that case we get an explicit expression for $Q_2$ in the limit of a slowly moving probe, taking $q_s = q_t$ in \eqref{Eq:DefQ1Q2} and we get
\bea
Q_2(t_1,t | q_t)  && =  e^{-r(t-t_1)}  \int_{0}^{+\infty} \int_{0}^{+\infty} r^2 d\tau d\tau' e^{-r(\tau +\tau')}  \left( A e^{- \frac{t-t_1 + \tau}{\gamma_b}} + q_t \frac{\gamma_b}{\gamma_{\lambda}}  \left(1 - e^{-\frac{t-t_1 + \tau}{\gamma_b}} \right)\right) \nn \\
&& \left(  A\left( e^{-r \tau} - e^{-r \tau'} \right) + \frac{\gamma_b}{\gamma_{\lambda}} q_t \left(  e^{-\frac{\min{\tau,\tau'}}{\gamma_b}} -e^{-\frac{\max{\tau,\tau'}}{\gamma_b}} \right) \right) \nn \, .
\eea
After some calculations we find
\bea  \label{Eq:ResultSecondCumulantConstantProbe}
 Q_2(t_1,t |q_t)  && = e^{-r_b(t-t_1)}  \left( A - q_t \frac{\gamma_b}{\gamma_{\lambda}} \right)  \left(A \frac{r \gamma _b}{4 r^2 \gamma _b^2+6 r \gamma _b+2} + q_t \frac{\gamma_b}{\gamma_{\lambda}}  \frac{r \gamma _b}{r^2 \gamma _b^2+3 r \gamma _b+2} \right) \nn \\
&& + 2 e^{-r(t-t_1)}  \left(  \frac{\gamma_b}{\gamma_{\lambda}} q_t \right)^2 \left( \frac{1}{-2 r \gamma _b-1}+\frac{1}{r \gamma _b+1} \right)  \, .
\eea

\section{Many-particles bath} \label{Sec:ManyParticleBath}

\subsection{Gaussian case}

In the case of a bath made off $N \gg 1$ particles, we consider the reduced equation for the probe dynamics \eqref{deo2} under the scaling limit \eqref{Eq:ScalingLimit} that we recall here:
\bea \label{Eq:StartingPointBath}
&& M \ddot{q_t} = -(a_1 +N a_3) q_t  + a_3  N \langle x_t^i | \{q_{t'}\}^t_{-\infty}\rangle + a_3 \sum_{i=1}^N \eta_t^i   \\
&& a_3 = a_3'/N \quad , \quad a_2 = a_2'/N \quad , \quad t = N t'  \quad , \quad  r =  r'/N  \quad , \quad M = N^2 M' \quad , \quad Q_{t'}= q_{Nt'}  \, . \nn 
\eea
Under this limit and using \eqref{Eq:FirstMomentExact} and \eqref{Eq:ExactResulKernel} (assuming $a_1'+a_2'+\gamma r'>0$) it is immediate to see that the first two terms on the right hand side of \eqref{Eq:StartingPointBath} converges as
\bea
\lim\limits_{N \to \infty}   -(a_1 +N a_3) q_t  + a_3  N \langle x_t^i | \{q_{t'}\}^t_{-\infty}\rangle  =  F_\text{tot}(Q_{t'}) - \int_{0}^{+\infty} \id\tau K'_{r'}(\tau) \dot{Q}_{t'-\tau}  \, ,
\eea
where $F_\text{tot}$ and the kernel $K'_{r'}$ were given in \eqref{Eq:IntroFtot} and \eqref{frk}. To evaluate the limit of the noise term $\sum_{i=1}^{N} \eta^i_t $, note that the noise terms are effectively uncorrelated when they are conditioned on a given probe history. It thus converges to a Gaussian noise:
\bea
\lim_{N \to \infty}   \sum_{i=1}^{N} \eta^i_{N t}  = \zeta(t')  \, ,
\eea
with $\zeta(t)$ a Gaussian noise with average $0$ and two-times correlation function given by (assuming $a_1'+a_2'+\gamma r'/2>0$), for $t_1' < t_1$
\bea
\langle \zeta(t') \zeta(t_1') \rangle  &=& \lim_{N \to \infty}    N \lambda^2 \langle  x_{Nt_1'} x_{Nt'} |\{q_{t'}\}_{-\infty}^{Nt'} \rangle \nn \\
& = & \lim_{N \to \infty}    N \lambda^2 \left( Q_1(N(t'-t_1')) + Q_2(Nt_1',Nt'| \{q_{-\infty}^{Nt_1'} \})   \right) \, ,
\eea
where we have used the decomposition \eqref{Eq:DefQ1Q2}. It is immediate to see that the first term converges as
\bea
\lim_{N \to \infty}    N \lambda^2  Q_1(N(t'-t_1'))  = T_\text{eff} \,K'_{r'}(t'-t_1') \, ,
\eea
in terms of the kernel \eqref{frk} and the effective temperature \eqref{Eq:IntroTeff}. On the other hand, it can be seen directly from the expression of $Q_2$ given in \eqref{Eq:DefQ1Q2} and of $\phi_1$ given in \eqref{Eq:DefPhi1Phi2} that the other contribution to the noise correlation becomes very small: $ N \lambda^2  Q_2(Nt_1',Nt'| \{q_{s}\}_{-\infty}^{Nt_1'}) = O(1/N^3)$. That follows from the fact that $\phi_1 = O(1/N)$ in this scaling limit. Hence we get that $\langle \zeta(t') \zeta(t_1') \rangle  = T_\text{eff} K'_{r'}(t'-t_1')$, reproducing the result given in \eqref{Eq:Intro:NoiseCorrelations}. The fact that $Q_2$ does not contribute in this limit explains why we can obtain an exact result in that case without the need of using a slow probe approximation.

\medskip

{\bf Remark} Note that the effective temperature interpretation only holds in the large $N$ limit. For small $N$ the noise is obviously nonGaussian but it also de-correlates on two different time scales (see \eqref{Eq:DefQ1Q2} and \eqref{Eq:ResultSecondCumulantConstantProbe}), while the friction kernel always (for finite $N$ as well) only involves the unique time-scale $1/(r+1/\gamma_b)$. 

\subsection{NonGaussian case}

We now discuss the rescaling to be performed in the nonGaussian case  $\gamma r \leq -2b$. In general the sum of iid distributed random variables $y_i$ distributed with a pdf exhibiting a fat tail $p(y) \sim_{y \to \pm \infty} |y|^{-1-\alpha}$ with $\alpha \leq 2$ converges to a generalized stable distribution if the random variables are rescaled as $N^{1/\alpha}$ (also substracting the mean if $\alpha \geq 1$). Here the noise felt by the probe $\zeta_t$ is $\zeta_t = \lambda \sum_{i=1}^N (x^i_t - \langle x^i_t | \{q_{t'}\}_{-\infty}^t \rangle)$ (without the average value substracted in the case $\gamma r \leq -b$), and the exponent of the fat-tail in the distribution of $x^i_t$ is $\alpha = - \frac{\gamma r}{b} $. Taking into account the prefactor in front of the tail (see \eqref{Eq:FatTail}) shows that  $x^i_t$ is of order $1/\sqrt{|b|}$ (the other constants are fixed). Hence we must take
\bea
\frac{\lambda}{\sqrt{|b|}} \sim \frac{1}{N^{1/\alpha}}   \, ,
\eea
which makes us choose $a_2 \sim  \frac{a_2'}{N^{2/\alpha}}$, $a_3 \sim  \frac{a_3'}{N^{2/\alpha}}$, $r \sim \frac{r'}{N^{2/\alpha}}$ (to conserve the value of the tail exponent), $t \sim t' N^{2/\alpha} $ (to keep $rt$ finite) and $M \sim M' N^{4/\alpha}$ (to keep $M \ddot{q}_t$ finite) as in Section \ref{ngc}.

In order to fully characterize the one-point distribution of $\zeta_t$ using a generalized central limit theorem (GCLT see \cite{GCLT}), we also need the prefactors of the tails of the distribution of bath particles given the probe history $p(x^i_t | \{ q_{t'} \}_{-\infty}^t )$ (that is not stationary). In principle these prefactors could depend on the full probe history but in the scaling limit considered here they do not depend on $q_t$. That can be seen directly on the expression of these prefactors that we gave in the case of a constant probe position: from \eqref{Eq:FatTail2} we obtain that the random variable $y_i = N^{1/\alpha}  \lambda x^i_t $ (with the average value substracted in the case $\alpha >1$) has a symmetric fat tail in the $N \to \infty$ limit: $p(y) \sim C/|y|^{1+\alpha} $ with 
\bea
C = \lim_{N \to \infty} \frac{1}{N^{1/\alpha} |\lambda|} G_{\pm} \frac{\gamma r/|b|}{(2T/|b|)^{\gamma r/(2b)}} (\lambda N^{1/\alpha})^{1+\alpha} = \alpha G \left( \frac{2 T a_3^2}{|a_2+a_3|}  \right)^{\alpha/2}  \, ,
\eea
where we recall that $G=  \frac{\Gamma(\frac{3+\alpha}{2})}{2\sqrt{\pi}}$. Here we have derived the result in the case where the probe stays at a constant position but it clear that this also holds in the general case. Using the GCLT of \cite{GCLT} we can conclude that the one time distribution of the noise is a stable distribution $S(\alpha , 0 , c , 0)$ with
\bea
c= \left(  \frac{\pi C}{\alpha \sin(\frac{\pi \alpha}{2}) \Gamma(\alpha))}  \right)^{1/\alpha}  \, ,
\eea
which leads to \eqref{Eq:StableDistc}. Here by definition $\zeta$ has a stable distribution $S(\alpha , 0 ,c , 0)$ if its characteristic function is $\langle e^{i \omega \zeta} \rangle = e^{ - c^{\alpha} |\omega|^{\alpha}} $.

{\bf Remark} The reason why the asymmetry in the tails of the distribution due to the non-zero value of the probe and of the resetting position does not appear in that scaling limit is because the temperature strongly dominates the short-time behavior of the evolution of the bath particles (which can be checked by rescaling the Langevin equation \eqref{Eq:1ParticleAndProbeLinearDynamics}). More precisely the bath particles feel the influence of the deterministic terms after a time of order $\gamma/|b|$, time at which they have diffused on a distance of order $\sqrt{D\gamma/|b|} = \sqrt{2T/|b|} \sim N^{1/\alpha}$. At that time the influence of the position of the resetting is lost since $A=O(1)$. Scaling also $A  \sim N^{1/\alpha}$, it can readily be seen from \eqref{Eq:FatTail2} that the asymmetry in the tail is conserved under the scaling limit, leading for the probe to a noise that is an asymmetric generalized stable distribution. A similar remark can be made for the Gaussian regime.

\section{Conclusion}
Adding random position resetting for overdamped bath particles with a linear dynamics yields an exactly solvable model for the probe evolution.  A Langevin dynamics can be derived in the large bath scaling limit and an interesting nonequilibrium effect arises: when the bath particles would run off to infinity under the equilibrium dynamics (without resetting) by the presence of an inverted harmonic well, the resetting stabilizes them and the result on the probe's effective dynamics is a nonGaussian noise, possibly with jumps having power-law tails. \\ 

\noindent {\bf Acknowledgment}: We are very grateful to Satya Majumdar for introducing us to the resetting dynamics and for useful suggestions on the paper. T.T. has been supported by the InterUniversity Attraction Pole phase VII/18 dynamics, geometry and statistical physics of the Belgian Science Policy.

\end{document}